\documentclass[usenatbib,usegraphicx]{mn2e}
\usepackage{amssymb}
\usepackage[fleqn]{amsmath}
\usepackage{charter}
\usepackage{mathdesign}
\usepackage{url}
\usepackage[hyperfootnotes, hyperfigures,bookmarks]{hyperref}
\hypersetup{bookmarksnumbered, bookmarksopen, colorlinks, citecolor=blue, pdfduplex=DuplexFlipLongEdge}
\usepackage{bookmark}
\providecommand{\eprint}[1]{\href{http://arxiv.org/abs/#1}{#1}}
\providecommand{\adsurl}[1]{\href{#1}{ADS}}
 
\usepackage{natbib}

\newcommand{\na} {New Astronomy}
\newcommand{\nat}   {Nature}
\newcommand{\pasj}  {PASJ}

\newcommand{\pre} {Phys. Rev. E}
\newcommand{\memsai} {Mem.~Soc.~Astron.~Italiana}

\usepackage{graphicx}
\usepackage{subfigure}

\usepackage{color}
\usepackage[usenames,dvipsnames]{xcolor}

\graphicspath{{figures/}}

\pagerange{\pageref{firstpage}--\pageref{lastpage}} \pubyear{2014}

\def\LaTeX{L\kern-.36em\raise.3ex\hbox{a}\kern-.15em
T\kern-.1667em\lower.7ex\hbox{E}\kern-.125emX}

\newcommand{\B}{\begin{eqnarray}}
\newcommand{\E}{\end{eqnarray}}

\newcommand{\bm}[1]{\mbox{\boldmath{$#1$}}}
\newcommand{\del}{{\bf \nabla}}

\title[Entropy Production in Relativistic Jet Boundary Layers]{Entropy Production in Relativistic Jet Boundary Layers}

\author[Kohler \& Begelman]
{Susanna Kohler$^{1, 2 \ast}$ and
Mitchell C. Begelman$^{1, 2 \star}$
\\$^1$ JILA, University of Colorado and National Institute of Standards and Technology, Boulder, CO 80309-0440, USA 
\\$^2$ Department of Astrophysical and Planetary Sciences, University of Colorado, Boulder, CO 80309-0391, USA
\\Email: $^\ast$ kohlers@colorado.edu,
$^\star$ mitch@jila.colorado.edu
}

\begin{document}

\label{firstpage}

\maketitle

\begin{abstract} 

Hot relativistic jets, passing through a background medium with a pressure gradient $p\propto r^{-\eta}$ where $2 < \eta \leq 8/3$, develop a shocked boundary layer containing a significant fraction of the jet power. In previous work, we developed a self-similar description of the boundary layer assuming isentropic flow, but we found that such models respect global energy conservation only for the special case $\eta = 8/3$. Here we demonstrate that models with $\eta < 8/3$ can be made self-consistent if we relax the assumption of constant specific entropy. Instead, the entropy must increase with increasing $r$ along the boundary layer, presumably due to multiple shocks driven into the flow as it gradually collimates.

The increase in specific entropy slows the acceleration rate of the flow and provides a source of internal energy that could be channeled into radiation. We suggest that this process may be important for determining the radiative characteristics of tidal disruption events and gamma-ray bursts from collapsars.

\end{abstract}


%%%%%%%%%%%%%%%%%%%%%%%%%%%%%%
%%%%%%%%%%%%%%%%%%%%%%%%%%%%%%
\section{Introduction} \label{intro}
%%%%%%%%%%%%%%%%%%%%%%%%%%%%%%
%%%%%%%%%%%%%%%%%%%%%%%%%%%%%%

Relativistic jets are becoming increasingly relevant as a component of high-energy astrophysical systems. Not only are these jets observed in the context of active galactic nuclei (AGN), microquasars, and gamma-ray bursts (GRBs), but they are now also considered to be an explanation of the flares seen from some tidal disruption events (TDEs), events in which a star is torn apart by the tidal forces exerted by a normally-dormant, massive black hole (\citealp{Zauderer11, Tchekhovskoy14}).

When a relativistic jet is launched, it slams into the gas and dust surrounding the source. Two things are thought to happen as these jets propagate outward. First, after they are launched with what are thought to be fairly modest speeds (\citealp{Georganopoulos98}), they are accelerated to Lorentz factors of typically a few for microquasars (\citealp{Meier03}), 10-20 for blazars (\citealp{Sikora94, Jorstad05}), and hundreds or even thousands for GRBs (\citealp{Lithwick01}). Second, they experience some form of collimation that results in the relatively narrow jet opening angles observed in most blazars and GRBs (e.g. \citealp{Doeleman12, Jorstad05, Sari99, Goldstein11}). As jets from active galactic nuclei are observed to first become collimated near their source (\citealp{JunorBiretta99, Jorstad05}), we seek a description to explain simultaneous collimation and acceleration at the base of the jet, where the internal-energy-dominated flow first interacts with the ambient medium.

Close to the central black hole, inside the Alfv\'en surface, jets can be collimated by the inertia of the disk at the jet base, transmitted by magnetic tension along the flow (e.g. \citealp{Blandford76, Lovelace76, Koide04, Narayan07}). At larger scales, however, causal connection between the disk and the flow is lost, creating a need for additional confinement by the pressure of an external medium (\citealp{MB95}). Additionally, collimation by magnetic tension has been demonstrated to occur very slowly (\citealp{Eichler93,Tomimatsu94,MB94,Beskin98}), and cannot explain the collimation scales observed.

In contrast to collimation where gas pressure is crucial, jet acceleration is generally assumed to be dominated by magnetic stresses. However, there are situations where other forces may play a dominant role. As an example, recent observations of TDEs indicate that some may exhibit powerful jets --- but there is not enough magnetic flux available to power such a jet without invoking a relic field (see e.g. \citealp{Tchekhovskoy14}). This provides additional motivation to study the effects of external pressure confinement on jet acceleration and collimation.

The environments around relativistic jets provide ideal scenarios for pressure confinement: there exist both static collimating environments such as the dusty torus of an AGN or the stellar envelope surrounding a GRB (e.g. \citealp{Eichler82, Komissarov97}), as well as the potential for collimation by dynamic means, such as by the ram pressure of a disk wind (e.g. \citealp{Komissarov94, BL07}). Pressure confinement and magnetic tension could work together to collimate a flow, as we describe in \citet{Kohler122}. Alternatively, pressure confinement could act alone --- a possibility which is our focus both in \citet{Kohler12}, hereafter KBB12, and in the current paper.

When a hydrodynamic, relativistic jet is injected into an ambient medium with a pressure profile that scales as $p \propto r^{-\eta}$ where $\eta>2$, the jet interior will ultimately lose causal contact with its surroundings (\citealp{MB84}). If $\eta~<~4$, a shocked boundary layer forms at the interface between the flow and the ambient medium, and the outer region of the jet experiences a collimating force from the pressure of that environment.

The range of $2<\eta<4$ for the external pressure profile could describe many scenarios for a confining medium, such as the ram pressure of a disk wind (see \citealp{Eichler82}), a disk corona, or a stellar envelope. This regime may be particularly relevant to the currently-explored topic of TDEs; in \citet{Coughlin14}, for example, the range calculated for the accretion disk that forms around the central black hole in a TDE is $3/2 < \eta <4$. Developing models of the jet flow in this range of ambient pressure profiles is therefore an important task for understanding the structure of the jets that form in a variety of interesting scenarios.

This problem has been previously approached in a variety of ways. Three-dimensional general relativistic MHD simulations such as \citet{Beckwith08a}, \citet{Beckwith09}, and \citet{McKinney09} provide self-consistent descriptions of the jet launching mechanism and propagation of a Poynting-flux-dominated jet sheathed by an unbound flow, but these simulations are limited by their inability to sufficiently resolve the sharp pressure and density gradients that occur in these regions. Other numerical studies of the large-scale collimation of jets treat the external medium as a rigid wall with a prescribed geometry, enclosing a cavity into which the jet is injected (e.g. \citealp{KomBarkVla07, KomVlaKon09, Komissarov11, Tchekhovskoy10}). Simulations that focus specifically on modeling gamma-ray burst jets breaking out of a stellar envelope --- which generally describe either Poynting-flux-dominated jets (e.g., \citealp{Proga03, Bromberg14}) or thermally-accelerated jets (e.g., \citealp{Aloy05, Lazzati05}) --- often suffer from similar constraints on resolution and boundary conditions, as well as the additional complications of time-dependent jet behavior (though the inclusion of this time-dependent analysis is of course more realistic than steady-state idealizations).

Because of these limitations, simplified analytic models are extremely helpful for improving the physical content of the boundary conditions employed in numerical simulations, as well as for interpreting the extent to which numerical resolution affects the outcomes of these simulations. As such, we opted to pursue an analytic approach that instead treats the external pressure as a boundary condition, but leave the physical shape of the boundary free to be determined as a result of interaction between the jet and the ambient medium.

Simplified analytic treatments of this problem can also vary greatly in approach, however. Works such as \citet{Komissarov97}, \citet{KN09}, and \citet{Lyubarsky09} focus on the interaction between a cold jet (dominated by inertia) and the ambient medium. In contrast, \citet{Levinson00} assumed a hot jet (dominated by the internal energy), but examined the structure and behavior of the jet in the limiting case where the jet becomes fully shocked upon impact. Other treatments, such as \citet{Zakamska08}, \citet{Begelman08}, and \citet{Lyubarsky11}, focus on modeling jets that remain in causal contact.

In contrast to these studies, we are interested in the case where the jet loses causal contact as it propagates, forming a boundary layer of shocked jet material at the point of impact with the ambient medium. This problem was previously addressed in \citet{BL07}; in this paper, the authors model the effects of both a stationary external medium and a disk wind on a hot, hydrodynamic jet that has lost causal contact when it impacts the external environment. The authors assume that the pressure remains constant across the boundary layer in this case, and they then examine the impact of the initial conditions, such as initial opening angle and Lorentz factor, on the structure of the jet as it balances its internal pressure with the pressure of the ambient medium.

In KBB12, we repeated their calculations, solving for the structure of a boundary layer just inside the contact discontinuity that separates the jet from its surroundings. We found results similar to those of \citet{BL07}, though we treated the entropy within the boundary layer slightly differently, which resulted in a difference in the collimating behavior of the outer boundary of the jet (see KBB12 for details). We then took our analysis a step further, however: due to the curvature of the jet as it collimates, we expect a pressure gradient to form across the boundary layer, so we opted to refine the boundary-layer treatment to allow for varying pressure across different streamlines. In the limit of ultrarelativistic flow (bulk Lorentz factor $\Gamma \gg 1$), we found a method of constructing self-similar models specifically for the structure within the boundary layer (see, again, KBB12 for results). These models, however, only gave physically reasonable results for certain values of $\eta$: global energy conservation was only respected for the special case where $\eta = 8/3$. Our models in KBB12 were also simplified by the use of several underlying assumptions that are commonly adopted in analytic treatments; in particular, that the flow in these solutions was both irrotational and isentropic.

In this paper, as a follow-up to KBB12, we now propose a set of more general solutions to the boundary-layer problem of a hot relativistic jet that has lost causal contact, and we demonstrate that these solutions provide physical results for the range of $\eta$ that was not well-described by KBB12 (which is the interesting regime for astrophysical scenarios such as TDEs). These solutions, unlike in our previous paper, do not require the jet to be either adiabatic or irrotational, and they allow for entropy to be a varying function of position within the boundary layer.

Our motivation in this work is, as before, to create steady-state models of the underlying jet behavior, onto which effects such as instabilities and radiation can be added. Our models provide insight into the locations of the bulk of the jet luminosity, as well as the locations of energy dissipation, which indicate potential radiation signatures that can be used for comparison with observations.

In \S \ref{physical} we describe the problem, summarize the behavior of the solutions we found in KBB12, and explain why we seek a new family of solutions here. In \S \ref{deriving} we describe how we obtain these new solutions, in \S \ref{results} we discuss the results and their implications, and in \S \ref{conclusion} we conclude.

%%%%%%%%%%%%%%%%%%%%%%%%%%%%%%
%%%%%%%%%%%%%%%%%%%%%%%%%%%%%%
\section{Casting the problem} \label{physical}
%%%%%%%%%%%%%%%%%%%%%%%%%%%%%%
%%%%%%%%%%%%%%%%%%%%%%%%%%%%%%

\begin{figure}
\center
\includegraphics[width=2.50in]{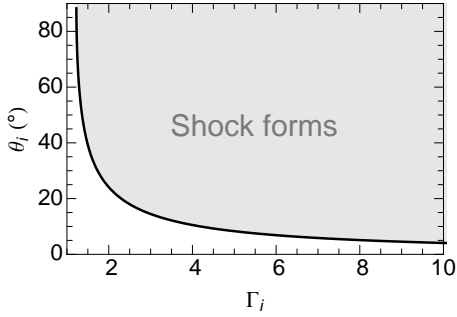}
\caption{Minimum angle of impact necessary to produce a shock, plotted as a function of the initial bulk Lorentz factor of the flow at the point of impact. The shaded region corresponds to combinations of angle and Lorentz factor that will result in a shock forming.}
\label{fig:minang}
\end{figure}

We assume an axisymmetric, ultrarelativistic, hot jet. Flow launched from the source initially expands adiabatically, propagating outward on conical streamlines until it impacts the ambient medium. Given a particular bulk Lorentz factor for the flow, there exists a minimum angle of impact that will result in the flow shocking. For the flow to shock, the speed of sound waves measured in the lab frame in the direction perpendicular to the shock surface must vanish; this is equivalent to the condition that $\beta_\perp= 1/\sqrt{2} \Gamma$, as described in KBB12. From this, and using the definition of the bulk Lorentz factor $\Gamma = (1-\beta^2)^{-1/2}$, one can show that the minimum angle of impact that will result in a shock is given by
	\B
	\sin{\theta_i} = \frac{1}{\sqrt{2} (\Gamma_i - 1)^{1/2}}, \label{minang}
	\E
where $\Gamma_i$ is the bulk Lorentz factor at the point of impact. Plotting $\theta_i$ in Figure \ref{fig:minang}, we can see that the angle of impact necessary to result in a shock is fairly small even for low values of $\Gamma_i$, indicating that formation of a shocked boundary layer is a very likely result.

% ===================================
% ===================================

%%%%%%%%%%%%%%%%%%%%%%%%%%%%%%
\subsection{Isentropic solutions} \label{isentropic}
%%%%%%%%%%%%%%%%%%%%%%%%%%%%%%

As mentioned in \S \ref{intro}, in KBB12 we sought self-similar solutions to the fluid equations within this shocked boundary layer. In this treatment we adopted several major simplifications to the problem: we assumed that the flow was adiabatic and isentropic, and that the flow was irrotational. In all solutions that we found, the overall shape of the jet remained very close to conical, but the relatively small amount of collimation due to the pressure of the ambient medium had a large impact on the behavior of the flow within the boundary layer.

The admitted solutions could be grouped into three families based upon how quickly the external pressure $p_e \propto r^{-\eta} $ dropped off. The behavior of each of those isentropic families is summarized briefly below.

\begin{enumerate}
	\item $\eta = 8/3$\\
	Solutions for which the external pressure profile is exactly $p_e \propto r^{-8/3}$ are a special case of solutions: the pressure within the boundary layer decreases from the contact discontinuity inward, dropping to zero at a finite distance and forming a narrow boundary layer containing a fixed flux of jet energy. These ``hollow cone'' solutions are the only isentropic solutions to conserve energy globally; the power carried by the jet, given by $L \propto p \Gamma^2 A \propto r^{-3 \eta / 4 +2}$ where $p$ is the pressure and $A$ is the cross-sectional area of the region carrying most of the power, is constant with radius in this case.
	\item $\eta > 8/3$\\
	Solutions for which the external pressure profile drops off more steeply than $p_e \propto r^{-8/3}$ have pressure within the boundary layer that decreases from the contact discontinuity inwards, but this pressure drops off asymptotically. The resulting solutions demonstrate ``hollow cone'' behavior for values of $\eta$ near $8/3$, but the boundary layer becomes wider and the cone progressively more filled as the value of $\eta$ increases toward 4. For these solutions, the power carried by the boundary layer scales as $r$ to a negative power. We demonstrated in KBB12, however, that because the transverse integrals of energy diverge, truncated solutions can still be constructed that conserve energy globally.
	\item $\eta < 8/3$\\
		There exist isentropic solutions for which the external pressure profile drops off less steeply than $p_e \propto r^{-8/3}$, however, for these solutions the power carried by the boundary layer scales as $r$ to a positive power. In KBB12, we argued that global energy could potentially be conserved if the jet cross-section gradually decreased with increasing radius, resulting in total energy remaining constant. We have since determined, however, that no such strongly-collimating solution can be constructed in a self-consistent way in the regime of $\eta < 8/3$. To find a set of solutions for the flow in this regime that \emph{do} behave physically, we must now relax some of the assumptions we made in KBB12 about the behavior of the flow.

\end{enumerate}

%%%%%%%%%%%%%%%%%%%%%%%%%%%%%%
\subsection{Entropy-generating solutions} \label{entropysolns}
%%%%%%%%%%%%%%%%%%%%%%%%%%%%%%

Why might the regime of $\eta < 8/3$ not have physical solutions with fixed entropy, as we assumed in KBB12? We argue here that our assumption of isentropic flow over-constrained the problem in this regime. This can be more readily understood from a physical point of view: consider a boundary layer in contact with an ambient medium that has a slowly-decreasing pressure profile. Because the pressure outside the jet is significantly higher than inside, and because the flow impacts the ambient medium at an oblique angle, the initial shock that forms at the point of impact may not decrease the speed of the flow enough to prevent it from shocking again further downstream.

\begin{figure}
\center
\includegraphics[width=3.25in]{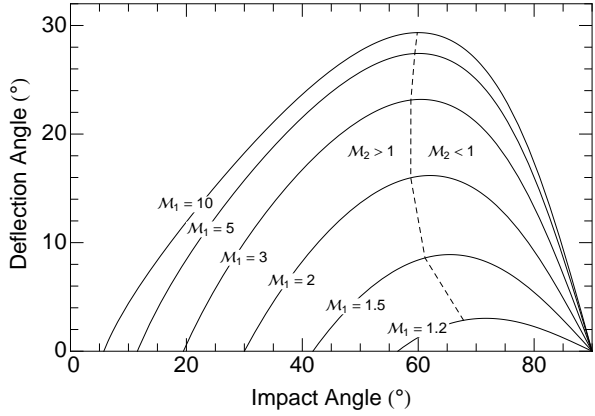}
\caption{Plot of the deflection angle of flow across a shock for a range of impact angles, assuming that the flow is supersonic upstream of the shock. Individual curves correspond to different values of the upstream Mach number, $\mathcal{M}_1 = \sqrt{2} (\Gamma_1^2 -1)^{1/2}$. The dashed line divides the parameter space into two regions: that in which the flow is subsonic downstream of the shock, and that in which it is still supersonic after shocking.}
\label{fig:maxangle}
\end{figure}

Figure \ref{fig:maxangle} illustrates, in the observer frame, the angle by which the flow is deflected when crossing a shock, as a function of the impact angle, for a series of curves describing different upstream Mach numbers. The plot also indicates the values of the deflection and impact angles for which the speed of the flow is still greater than the sound speed, downstream of the shock; from this, it can be seen that for an oblique shock the flow can easily remain supersonic after shocking. In addition, we will later demonstrate that the flow continues to be accelerated after crossing to the downstream side of the shock. This acceleration makes it even more likely that the flow could shock again as it propagates outward in radius. One can imagine, then, the jet flow undergoing not just one, but a series of shocks, with each successively decreasing the speed perpendicular to the shock and deflecting streamlines closer to the shock tangent. This process would result in a gradually collimating flow, but each shock would also generate entropy, providing a source of internal energy.

Flow undergoing a series of shocks in this way would be difficult to model analytically; however, the motion can be approximated. By envisioning the flow as forming a single shocked boundary layer --- but allowing the specific entropy within the layer to vary spatially --- we can approximate the entropy generation that would occur over the course of a series of shocks. By not insisting that the fluid be irrotational, we provide the additional freedom needed for the flow to mimic the behavior of crossing multiple shocks.

Using this physical model, we now describe the construction of the problem under these terms and the solutions that are admitted as a result.

% ===================================
% ===================================

%%%%%%%%%%%%%%%%%%%%%%%%%%%%%%
%%%%%%%%%%%%%%%%%%%%%%%%%%%%%%
\section{Deriving the boundary-layer flow} \label{deriving}
%%%%%%%%%%%%%%%%%%%%%%%%%%%%%%
%%%%%%%%%%%%%%%%%%%%%%%%%%%%%%

As in KBB12, we model a hot, ultrarelativistic jet that is cylindrically symmetric about the $z$ axis. We assume that the jet is injected from a point source with steady flow and approximately radial streamlines, and we suppose that the ambient medium has a pressure profile that declines as $p_e \propto r^{- \eta}$ where $r$ is spherical radius and $\eta$ is some constant satisfying $2 < \eta < 4$ (see \citealp{BL07} for another treatment of this regime). For an external pressure that decreases at such a rate, the jet will lose causal contact and become gradually collimated by the ambient pressure; we now examine, in the steady-state limit, the shape that the jet takes as a result of this collimation.

When the jet plows into the external medium supersonically, a boundary layer of shocked jet material forms at the interface between the jet and the stationary ambient medium. We model this boundary layer with a thickness of order $\Delta \theta \sim 1/ \Gamma$ such that the layer maintains causal contact, and we wrap the physics of the corresponding ambient medium into the pressure profile $p_e$.

In our model, the initial opening angle of the jet is assumed to be less than $\pi/2$ and greater than the minimum angle of impact for a shock, given by Eq~\eqref{minang}, such that causal contact is lost across the jet and a shock forms when the jet impacts the ambient medium. We note that the initial opening angle does not otherwise play a role in establishing the structure that is formed across the boundary layer, because we construct here solutions of the boundary layer only, rather than attempting to solve for the entire structure of the jet. For further insight into the full jet structure, as well as details about how the initial conditions at the jet base affect the jet shape and behavior, we refer the reader to our previous paper, KBB12, as well as other works (e.g. \citealp{Levinson00, BL07}).

%%%%%%%%%%%%%%%%%%%%%%%%%%%%%%
\subsection{Fluid equations} \label{fluideqs}
%%%%%%%%%%%%%%%%%%%%%%%%%%%%%%

The boundary layer of the jet is bounded on the inside by a shock front, through which jet material enters, and on the outside by a contact discontinuity, separating it from the ambient medium. To find solutions describing the fluid flow within the boundary layer, we start from the covariant form of the relativistic, hydrodynamic fluid equations (e.g. \citealp{Dixon78,Zakamska08}):
	\B
	\del_\nu (\rho u^\nu) &=& 0\\
	\del_\nu T^{\mu \nu} &=& 0,
	\E
where
	\B
	T^{\mu \nu} = w u^\mu u^\nu + p g^{\mu \nu}
	\E
is the stress-energy tensor. Here $\rho$ is the proper rest mass density, $w$ is the enthalpy, $p$ is the pressure, $u^{\mu} = (\Gamma, \Gamma \bm \beta)$ is the 4-velocity of a fluid element (with $\bm \beta = \bm v / c$), and $g^{\mu \nu}$ is the space metric. The enthalpy is defined as $w \equiv \epsilon + p$ where $\epsilon$ is the total proper energy density, given by $\epsilon = \rho + 3 p$.

Assuming flat spacetime and time independence, and using number density $n$, these equations can be written in vector notation as
	\B
	\del \cdot (n \Gamma \bm \beta) &=& 0 \label{cont} \\
	\del \cdot (w \Gamma^2 \bm \beta) &=& 0 \label{eneq} \\
	w \Gamma^2 (\bm \beta \cdot \del) \bm \beta + \del p &=& 0, \label{momeq}
	\E	
describing continuity, energy conservation, and momentum conservation, respectively.

Treating the problem in the spherical polar coordinates $r, \theta$ and $\phi$, we examine the ordering of the various terms in these equations. As the maximum transverse speed that can be achieved without a shock forming is of order ${1}/{\Gamma}$, we can assume that $\beta_\theta$ is of this order, and that $\beta_r$ is of order one. Adopting this characteristic scaling, we state that $\frac{\partial}{\partial \theta} \sim \Gamma \frac{\partial}{\partial r} $. Writing out $\beta_r$ and employing the fact that $\beta_\theta^2 + \Gamma^{-2} \ll 1$, we have $\beta_r \approx 1- \frac{1}{2} ( \beta_\theta^2 + \Gamma^{-2})$. Finally, we also assume that $\theta \approx$ constant, i.e., that the streamlines are roughly radial; we will calculate their deviation from radial. Using all of these arguments, we hereafter retain terms only to lowest order.

%%%%%%%%%%%%%%%%%%%%%%%%%%%%%%
\subsection{Constructing self-similar solutions} \label{ss}
%%%%%%%%%%%%%%%%%%%%%%%%%%%%%%

As in KBB12, we construct a self-similar variable $\xi$ that describes the distance into the boundary layer from the contact discontinuity, normalized by the expected scale of the boundary layer: $\xi \propto (\theta_c - \theta)/{\Delta \theta}$, where $\theta_c = \theta_c(r)$ is the position of the contact discontinuity. Assuming that $\rho \ll p$ such that $w \approx 4p$, we now search for solutions to the fluid equations that take the form
	\begin{align}
	p &= g^4(\xi) r^{-\eta},&  \quad \beta_\theta &= h(\xi) r^{-\delta}, \nonumber \\
	\dfrac{1}{\Gamma} &= j(\xi) r^{-\delta}, &  \quad  n &= k(\xi) r^{- \alpha}. \label{solns}
	\end{align}
In these solutions $\delta$ and $\alpha$ are constant free parameters describing radial scaling, and $g$, $h$, $j$ and $k$ are arbitrary functions of the similarity variable $\xi$, with notation chosen for consistency with KBB12. The interior pressure has the same radial scaling as the pressure of the ambient medium because the gas pressure in the boundary layer must match the external pressure at the contact discontinuity. The boundary condition $g(0)=1$ is enforced so that the pressures are matched at the contact discontinuity, but $h(0)$ is left free. Finally, the streamlines must be parallel to the contact discontinuity at its location, $\theta_c$, which requires that $\beta_\theta(\theta_c) = r d\theta_c / dr$ and yields the further constraint that
	\B
	\dfrac{d\theta_c}{dr} &=& h(0) r^{-(1+\delta)}.
	\E

In contrast to KBB12, where we did not specifically enforce that the boundary layer maintain constant power, we now impose this condition. The power carried by a pressure-dominated jet is given by $L \propto p \Gamma^2 A$, where $A$ is the cross-sectional area of the jet. If we examine the jet's power concentrated in the boundary layer, which has width $\Delta \theta \sim 1/ \Gamma$, then $L \propto p \Gamma r^2$. Using the pressure scaling $p \propto r^{-\eta}$ we see that for power to remain constant, the radial scaling of the Lorentz factor within the layer must be $\delta = {\eta-2}$.

By taking the scalar and vector products of $\bm \beta$ with Eq~\eqref{momeq}, the momentum equation can be broken down into components parallel and perpendicular to the fluid motion, respectively. It can be shown (e.g., \citealp{Landau59}) that the parallel component of the momentum conservation equation can be reworked to obtain the relativistic Bernoulli equation:
	\B
	\frac{\Gamma w}{n} = B \label{bernoulli}
	\E
where $B$ is a constant function along streamlines.

Unlike in KBB12, we make no assumptions that the flow is either irrotational or adiabatic. Instead, we obtain an additional equation by supposing that there is no flow across surfaces of constant $\xi$; that is, we claim that $\xi$ is a streamfunction. Thus, to Eqs \eqref{cont} -- \eqref{momeq} we add one final equation:
	\B
	\bm \beta \cdot \del \xi = 0. \label{streamlines}
	\E

This definition clarifies that the Bernoulli constant $B=B(\xi)$. Thus Eq \eqref{bernoulli} allows us to determine both the radial scaling for $n$ and the transverse distribution for $\Gamma$: $\alpha = 2$ and $j(\xi)~=~\frac{4 g^4(\xi)}{B(\xi) k(\xi)}$.

As the boundary layer is expected to scale as $\Delta \theta = {1}/{\Gamma_c}$, where $\Gamma_c$ is the Lorenz factor along the contact discontinuity, we can define our self-similar variable $\xi$ as
	\B
	\xi = - \dfrac{1}{h(0)} r^{\eta - 2} (\theta_c - \theta).
	\E

Using this definition with Eq \eqref{streamlines}, we can solve for the form of the spatial distribution of the transverse velocity, $h(\xi)$. Thus we now have
	\begin{align}
	p &= g^4(\xi) r^{-\eta},&  \quad \beta_\theta &= h(0)\left(1 - (\eta - 2)\xi \right) r^{2-\eta}, \nonumber \\
	\dfrac{1}{\Gamma} &= \frac{4 g^4(\xi)}{B(\xi) k(\xi)} r^{2-\eta}, &  \quad  n &= k(\xi) r^{-2}. \label{solns2}
	\end{align}
Note here that $h(0)$ is a negative value, in order to provide collimating solutions for $\beta_\theta$.

Finally, turning to Eq \eqref{momeq} and keeping terms only to lowest order, we can write the perpendicular component of the momentum conservation equation as
	\B
	w \Gamma^2 \left( \frac{\partial \beta_\theta}{\partial r} + \frac{\beta_\theta}{r} \frac{\partial \beta_\theta}{\partial \theta} + \frac{\beta_\theta}{r}  \right) + \frac{1}{r} \frac{\partial p}{\partial \theta} = 0.
	\E
Substituting the solutions from Eq \eqref{solns2} yields a differential equation that governs the behavior of $g(\xi)$ and $k(\xi)$:
	\B
	\left( \frac{B(\xi) k(\xi) h(0)}{4 g^4(\xi)} \right)^2 \left( 3 - \eta \right) \left(1 - (\eta - 2) \xi \right) + \frac{g'(\xi)}{g(\xi)} = 0, \label{de}
	\E
where the prime indicates a derivative with respect to $\xi$. We now explore these results further.

%%%%%%%%%%%%%%%%%%%%%%%%%%%%%%
%%%%%%%%%%%%%%%%%%%%%%%%%%%%%%
\section{Results} \label{results}
%%%%%%%%%%%%%%%%%%%%%%%%%%%%%%
%%%%%%%%%%%%%%%%%%%%%%%%%%%%%%

%%%%%%%%%%%%%%%%%%%%%%%%%%%%%%
\subsection{Radial scaling}
%%%%%%%%%%%%%%%%%%%%%%%%%%%%%%

Looking at the form of the solutions to Eq \eqref{solns2}, we can first discuss the evident radial scalings. By construction, the pressure within the layer decreases radially with the same scaling as the pressure of the ambient medium. In the entropy-generating solutions for $\eta < 8/3$ here, the Lorentz factor must scale as $\Gamma \propto r^{\eta-2}$ in order to satisfy energy conservation within the layer, whereas in the isentropic solutions found for $\eta \ge 8/3$ in KBB12, the Bernoulli equation instead forced the Lorentz factor to scale as $\Gamma \propto r^{\eta/4}$.

\begin{figure}
\center
\includegraphics[width=3.25in]{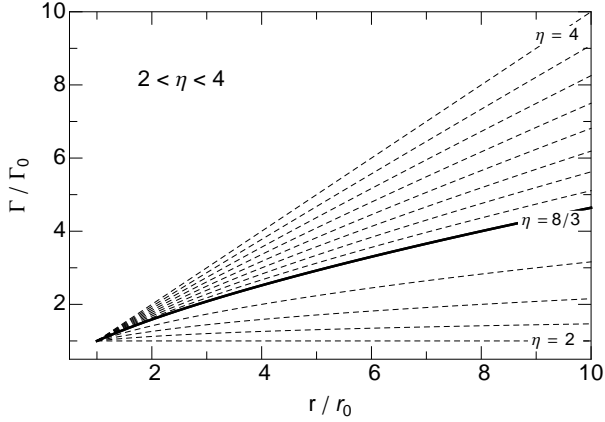}
\caption{Radial scaling of the bulk Lorentz factor for curves with differing values of the pressure power-law index $\eta$, ranging from $\eta = 2$ to $\eta = 4$  in increments of 1/6.}
\label{fig:radialgamma}
\end{figure}

Figure \ref{fig:radialgamma} illustrates the scaling of $\Gamma$ for curves in both of these ranges: the thick line corresponds to $\Gamma(r)$ for $\eta = 8/3$, the curves below correspond to entropy-generating solutions with values of $\eta$ less than 8/3, and the curves above correspond to isentropic solutions with values of $\eta$ greater than 8/3. From this plot, we can see that the process of entropy generation in the range of $\eta < 8/3$ slows the acceleration rate of the flow as compared to the constant-entropy solutions in the range of $\eta \ge 8/3$. Nonetheless, in both ranges the Lorentz factor scales always as a positive power of radius, verifying that flow accelerates within the boundary layer as it travels outward in radius. This result supports our picture of the flow described in \S \ref{entropysolns}, wherein the fluid undergoes repeated shocking as a result of being reaccelerated to supersonic velocities.

% ===================================
% ===================================

Examining other radial scalings in the $\eta < 8/3$ solutions, we see that the transverse velocity decreases as the fluid travels further from the source, and the density also decreases, as expected. Note that the radial scaling for density here is $n \propto r^{-2}$, which is consistent with an adiabatic equation of state (wherein $p~\propto~n^{4/3}$) only in the special case where $\eta=8/3$.

%%%%%%%%%%%%%%%%%%%%%%%%%%%%%%
\subsection{Transverse scaling: a linear family of solutions}
%%%%%%%%%%%%%%%%%%%%%%%%%%%%%%

The end equations from \S \ref{ss} admit a wide range of possible solutions, and thus flow can be constructed using an appropriate combination of density profile and varying Bernoulli parameter to fit the physical circumstances of a given system. For the purpose of examining the solutions further, however, let us consider a simple set of solutions closely related to those we explored in KBB12: the solutions that arise when $g(\xi)$ is linear.

%%%%%%%%%%%%%%%%%%%%%%%%%%%%%%
\subsubsection{General linear solutions} \label{gensolns}
%%%%%%%%%%%%%%%%%%%%%%%%%%%%%%

Linear solutions are already hinted at in the form that $\beta_\theta$ is demonstrated to take in Eq \eqref{solns2}; furthermore, examining linear solutions is the simplest approach and will allow us also to draw analogies to the solutions in KBB12. So let us suppose that
	\B
	g(\xi) = 1 - s \xi \label{glinear}
	\E
where $s$ is some constant greater than zero. Using this form, Eq \eqref{de} reduces to
	\B
	B(\xi) k(\xi) = - \frac{4}{h(0)} \left( \frac{s}{3-\eta} \right)^{1/2} \left( \frac{(1-s \xi)^7}{1-(\eta-2) \xi} \right)^{1/2}. \label{linear}
	\E
We use the negative root from Eq \eqref{de} here because $h(0)$ is negative and from the expression for $\Gamma$ in Eq \eqref{solns2}, we see that $B(\xi)k(\xi)$ must be positive to obtain physical results. From here we can now examine the solutions that arise:
	\B
	p &=& \left(1-s \xi \right)^4 r^{-\eta} \nonumber \\
	\beta_\theta &=& h(0)\left(1 - (\eta - 2)\xi \right) r^{2-\eta} \nonumber \\
	\dfrac{1}{\Gamma} &=& - h(0) \left(\frac{3- \eta}{s} \right)^{1/2} \left(1-s \xi \right)^{1/2} \left(1-(\eta-2)\xi \right)^{1/2} r^{2-\eta} \nonumber\\
	n &=& k(\xi) r^{-2} \nonumber \\
	&=& - \frac{4}{h(0)} \left( \frac{\eta-2}{3-\eta} \right)^{1/2} \frac{\left(1-(\eta-2) \xi \right)^{3}}{B(\xi)} r^{-2}. \label{solnslin}
	\E
A few things are evident looking at these solutions. First, the pressure, transverse velocity, and inverse of the Lorentz factor all clearly drop to zero at finite values of $\xi$ (as does the density, depending on the function chosen for the Bernoulli parameter). Thus the boundary layer that forms is either of width $\Delta \xi= 1/s$ or $\Delta \xi= 1/(\eta-2)$ --- whichever is smaller. This is a ``hollow cone'' solution for the jet, as described in \S \ref{isentropic}; the jet has a narrow boundary layer that contains a fixed energy flux.

The form of the density profile in this set of solutions remains free; it can be specified either by fixing the profile itself by prescribing $k(\xi)$, or by selecting a physically-motivated function to describe how the Bernoulli parameter $B(\xi)$ evolves. This latter point will be discussed further in \S \ref{entropygen}.

%%%%%%%%%%%%%%%%%%%%%%%%%%%%%%
\subsubsection{Linear solutions when $s=\eta-2$}
%%%%%%%%%%%%%%%%%%%%%%%%%%%%%%

We can further examine a specific example of this set of solutions: that in which $s=\eta-2$. In this case, the solutions are given by
	\B
	p &=& \left(1-(\eta-2) \xi \right)^4 r^{-\eta} \nonumber \\
	\beta_\theta &=& h(0)\left(1 - (\eta - 2)\xi \right) r^{2-\eta} \nonumber \\
	\dfrac{1}{\Gamma} &=& - h(0) \left(\frac{3- \eta}{\eta-2} \right)^{1/2} \left(1-(\eta-2)\xi \right) r^{2-\eta} \nonumber\\
	n &=& k(\xi) r^{-2}. \label{solnslinsfixed}
	\E

We can see that this is a variable-entropy generalization of the solution found in KBB12 for the special case of $\eta = 8/3$; that solution can be reproduced by assuming constant $B(\xi)$, in which case $p \propto n^{4/3}$. The generalized solution here behaves similarly: the flow forms a narrow boundary layer wherein the pressure drops from the value of the external pressure at $\xi=0$ (the contact discontinuity) to zero at $\xi = 3/2$, implying that the jet material is piled up in a thin sheath around the outside of the flow.

When we instead examine a general $\eta$, the pressure and transverse velocity still drop to zero at a finite value of $\xi$, but the width of the boundary layer that forms is $\eta$-dependent: $p$ and $\beta_\theta$ both go to zero when $\xi = 1/(\eta-2)$. Thus the boundary layer still has a finite width, but that width increases as $\eta$ decreases from $\eta = 8/3$ to $\eta \rightarrow 2$, and the hollow cone structure widens.

\begin{figure}
\center

\subfigure[]{
	\includegraphics [width=3.3in] {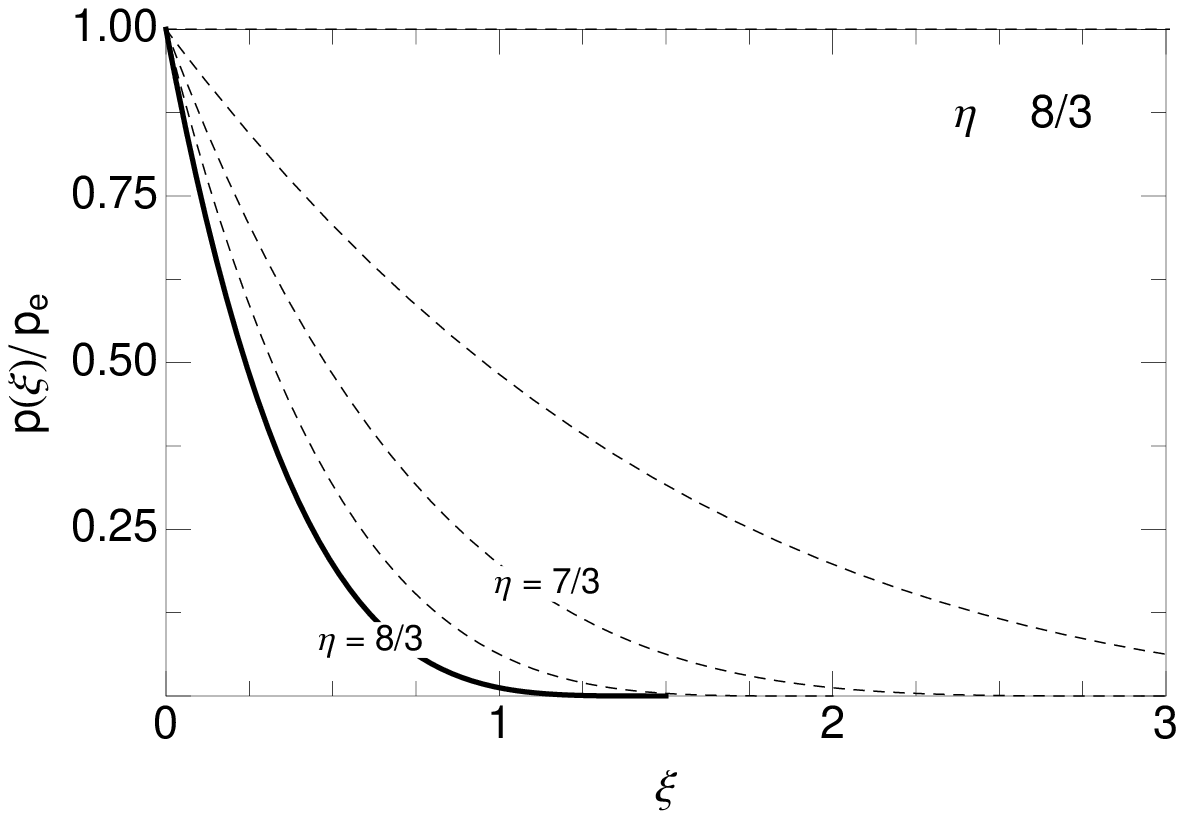}
	\label{fig:press1}
	}
\subfigure[]{
	\includegraphics [width=3.3in] {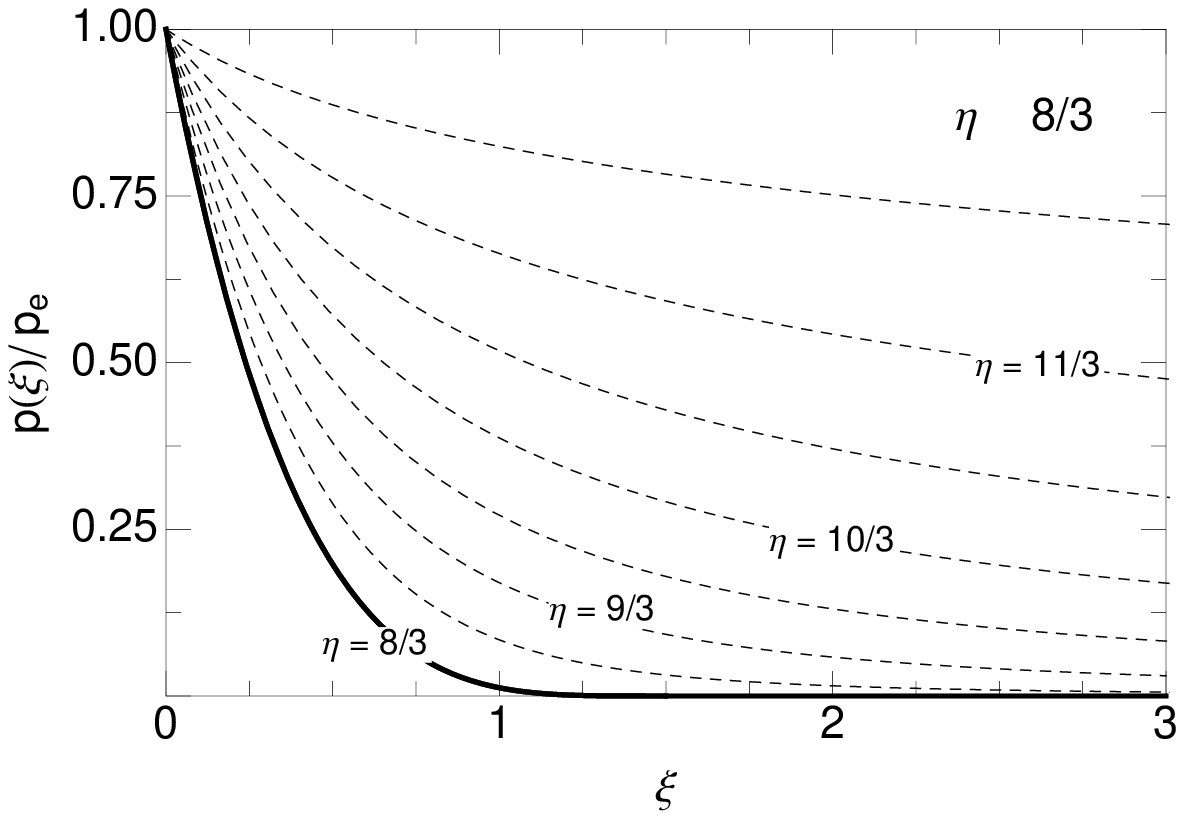}
	\label{fig:press2}
	}
\caption{Pressure within the shock as scaled by the external pressure, for curves of constant $\eta$. The contact discontinuity is located at $\xi=0$. (a)~Curves for $2 < \eta \le 8/3$, with $\eta$ decreasing from the bottom curve up in increments of $1/6$. In these solutions, pressure drops to zero at a finite value of $\xi$ for each curve. (b)~For comparison (from KBB12), curves for $8/3 \le \eta < 4$, with $\eta$ decreasing from the bottom curve up in increments of $1/6$. In these solutions, pressure decreases asymptotically.}
\label{fig:press}
\end{figure}

We can compare the transverse distribution of pressure in this particular case to that found in KBB12 for $\eta \ge 8/3$. Figure \ref{fig:press} illustrates this difference, showing how much more steeply the pressure drops off in the solutions where $\eta < 8/3$ (Figure \ref{fig:press1}) than in the solutions where $\eta \ge 8/3$ (Figure \ref{fig:press2}).

% ===================================
% ===================================

%%%%%%%%%%%%%%%%%%%%%%%%%%%%%%
\subsection{Entropy generation} \label{entropygen}
%%%%%%%%%%%%%%%%%%%%%%%%%%%%%%

We can now calculate the specific entropy within the boundary layer using the solutions derived in \S \ref{deriving}, and see if this is consistent with the assumptions we made in constructing the problem, as described in \S \ref{entropysolns}.

The specific entropy within the boundary layer scales as
	\B
	\sigma \propto \ln \left(\frac{p}{n^\gamma}\right)
	\E
where the adiabatic index is $\gamma=4/3$ in our case of a relativistic jet. In the solutions proposed in this paper, $n \propto r^{-2}$, meaning that the specific entropy scales as $\sigma \propto \ln (r^{8/3 - \eta})$. This is a constant for $\eta = 8/3$, which is consistent with the constant-entropy solutions we found in KBB12. For $\eta < 8/3$, the entropy increases with $r$, which is indeed consistent with the theory that the fluid undergoes a series of entropy-generating shocks when the ambient pressure decreases slowly. Finally, for $\eta > 8/3$, entropy is a decreasing function of $r$, indicating that these solutions don't make sense in this regime.

Given the different behavior of these three different regimes, it is therefore important to select the free parameters for the solutions in a way that is consistent with the physical picture. The Bernoulli function $B(\xi)$ is an initial condition, set by the description of the interior jet flow as it crosses the shock front and enters the boundary layer; $B(\xi)$ remains the same on a streamline, even when crossing a shock. Pressure and density profiles should therefore be selected carefully, both to ensure consistency with the initial flow, and to ensure that the second law of thermodynamics isn't violated.

%%%%%%%%%%%%%%%%%%%%%%%%%%%%%%
\subsection{Additional Isentropic Solutions}
%%%%%%%%%%%%%%%%%%%%%%%%%%%%%%

Interestingly, we can also find solutions for $\eta > 8/3$ using the approach described in \S \ref{deriving}; the primary difference between these and the $\eta < 8/3$ solutions is that here we can't require energy to be conserved. The resulting solutions for $\eta > 8/3$ take the form
	\begin{align}
	p &= g^4(\xi) r^{-\eta},&  \quad \beta_\theta &= h(0)\left(1 - \dfrac{\eta}{4} \xi \right) r^{-\eta / 4}, \nonumber \\
	\dfrac{1}{\Gamma} &= \frac{4 g^4(\xi)}{B(\xi) k(\xi)} r^{- \eta / 4}, &  \quad  n &= k(\xi) r^{-3 \eta / 4}. \label{greatersolns}
	\end{align}
In this case the flow is isentropic, as in KBB12, because the collimation is too weak to generate multiple shocks. In contrast to KBB12, however, these solutions are not irrotational, and their density profiles are not prescribed.

For these solutions, the energy flux within the boundary layer decreases with increasing radial distance from the source. To create physical solutions, forms of the density and pressure profiles can be chosen that diverge in the transverse direction, as in KBB12, allowing for constant total energy flux within the layer when the bounding shock moves to larger values of $\xi$ with increasing $r$. Linear solutions like those described in \S \ref{gensolns}, however, are possible; they merely require that the inner boundary is not in fact a shock front. Instead, this boundary must be something akin to a rarefaction front, through which material leaves the boundary layer and rejoins the main jet flow.

For an ambient medium with a pressure profile where $\eta > 8/3$, then, both the irrotational family of solutions found in KBB12 and the family of solutions described above are isentropic, and are reasonable physical descriptions of the flow in this regime.

%%%%%%%%%%%%%%%%%%%%%%%%%%%%%%
%%%%%%%%%%%%%%%%%%%%%%%%%%%%%%
\section{Conclusion} \label{conclusion}
%%%%%%%%%%%%%%%%%%%%%%%%%%%%%%
%%%%%%%%%%%%%%%%%%%%%%%%%%%%%%

Our goal was to obtain solutions that describe boundary-layer flow in a hot, relativistic jet that has lost causal contact with the ambient medium. We first sought solutions that specifically ensured that energy is conserved within the layer. We confirm here that for an external pressure profile of $p_e \propto r^{-\eta}$, when $\eta = 8/3$, linear solutions are possible in which the jet forms a narrow boundary layer of width $\Delta \xi = 1.5$ and behaves as a hollow cone, as described in KBB12, with all of the jet material piled up against the outer edge. This solution conserves energy within the boundary layer and the flow is irrotational, isentropic, and behaves adiabatically.

For a more gradual drop in pressure, where $\eta < 8/3$, the treatment in KBB12 was not sufficient; instead, the solutions demonstrated in this paper are the physical ones. In these solutions, we ensure conservation of energy within the boundary layer, but relax the conditions that the flow must be isentropic and irrotational. As a result, we find solutions in which the flow undergoes repeated shocking, generating entropy as it propagates outward in radius and gradually collimating in the process.

Linear solutions can be found in this regime that behave somewhat similarly to those for $\eta = 8/3$: a boundary layer forms in which the pressure drops to zero at a finite point, prescribing the width of the layer. Here, however, the layer width is dependent upon how steeply the external pressure decreases: as $\eta$ decreases, the layer becomes broader, and it will continue to widen until the cone of the jet is filled. Furthermore, the rate of acceleration is slowed by the entropy generation, and the flow must no longer behave adiabatically --- in fact, the density profile in this regime is completely free, and can be chosen to fit the physical parameters of a given scenario.

Finally, for a steeper drop in external pressure, where $\eta > 8/3$, we find a new family of solutions that are isentropic, as in KBB12, but for which the flow is not irrotational and the density profile is left free. Linear forms of these solutions result in an energy flux that decreases radially, suggesting that the inner edge of the boundary layer may be a rarefaction front in this case.

This freedom to arbitrarily set up the system allows us to apply this model to a variety of different astrophysical flows; as previously mentioned, any hot, relativistic jets without strong magnetic fields could conceivably be described by this model. The entropy generation included in this model could provide a direct source of internal energy leading to radiation; thus this model is an important first step in establishing the radiation characteristics of these astrophysical flows.

%%%%%%%%%%%%%%%%%%%%%%%%%%%%%%
%%%%%%%%%%%%%%%%%%%%%%%%%%%%%%
\section*{Acknowledgements}
%%%%%%%%%%%%%%%%%%%%%%%%%%%%%%
%%%%%%%%%%%%%%%%%%%%%%%%%%%%%%

This work was supported in part by NSF grant AST-0907872, NASA Astrophysics Theory Program grant NNX09AG02G, and NASA's Fermi Gamma-ray Space Telescope Guest Investigator program.

%\newpage

%\bibliographystyle{hapj}

%\bibliography{refsthesis}

\begin{thebibliography}{}

\bibitem[\protect\citeauthoryear{{Aloy}, {Janka} \& {M{\"u}ller}}{{Aloy}
  et~al.}{2005}]{Aloy05}
{Aloy} M.~A.,  {Janka} H.-T.,    {M{\"u}ller} E.,  2005, A\&A, 436, 273,
  \eprint{astro-ph/0408291},
  \adsurl{http://adsabs.harvard.edu/abs/2005A\%26A...436..273A}

\bibitem[\protect\citeauthoryear{{Beckwith}, {Hawley} \& {Krolik}}{{Beckwith}
  et~al.}{2008}]{Beckwith08a}
{Beckwith} K.,  {Hawley} J.~F.,    {Krolik} J.~H.,  2008, ApJ, 678, 1180,
  \eprint{0709.3833},
  \adsurl{http://adsabs.harvard.edu/abs/2008ApJ...678.1180B}

\bibitem[\protect\citeauthoryear{{Beckwith}, {Hawley} \& {Krolik}}{{Beckwith}
  et~al.}{2009}]{Beckwith09}
{Beckwith} K.,  {Hawley} J.~F.,    {Krolik} J.~H.,  2009, ApJ, 707, 428,
  \eprint{0906.2784},
  \adsurl{http://adsabs.harvard.edu/abs/2009ApJ...707..428B}

\bibitem[\protect\citeauthoryear{{Begelman}}{{Begelman}}{1995}]{MB95}
{Begelman} M.~C.,  1995, Proceedings of the National Academy of Science, 921,
  11442, \adsurl{http://adsabs.harvard.edu/abs/1995PNAS...9211442B}

\bibitem[\protect\citeauthoryear{{Begelman}, {Blandford} \& {Rees}}{{Begelman}
  et~al.}{1984}]{MB84}
{Begelman} M.~C.,  {Blandford} R.~D.,    {Rees} M.~J.,  1984, Reviews of Modern
  Physics, 56, 255, \adsurl{http://adsabs.harvard.edu/abs/1984RvMP...56..255B}

\bibitem[\protect\citeauthoryear{{Begelman}, {Fabian} \& {Rees}}{{Begelman}
  et~al.}{2008}]{Begelman08}
{Begelman} M.~C.,  {Fabian} A.~C.,    {Rees} M.~J.,  2008, MNRAS, 384, L19,
  \eprint{0709.0540},
  \adsurl{http://adsabs.harvard.edu/abs/2008MNRAS.384L..19B}

\bibitem[\protect\citeauthoryear{{Begelman} \& {Li}}{{Begelman} \&
  {Li}}{1994}]{MB94}
{Begelman} M.~C.,  {Li} Z.-Y.,  1994, ApJ, 426, 269,
  \adsurl{http://adsabs.harvard.edu/abs/1994ApJ...426..269B}

\bibitem[\protect\citeauthoryear{{Beskin}, {Kuznetsova} \& {Rafikov}}{{Beskin}
  et~al.}{1998}]{Beskin98}
{Beskin} V.~S.,  {Kuznetsova} I.~V.,    {Rafikov} R.~R.,  1998, MNRAS, 299,
  341, \adsurl{http://adsabs.harvard.edu/abs/1998MNRAS.299..341B}

\bibitem[\protect\citeauthoryear{{Blandford}}{{Blandford}}{1976}]{Blandford76}
{Blandford} R.~D.,  1976, MNRAS, 176, 465,
  \adsurl{http://adsabs.harvard.edu/abs/1976MNRAS.176..465B}

\bibitem[\protect\citeauthoryear{{Bromberg}, {Granot}, {Lyubarsky} \&
  {Piran}}{{Bromberg} et~al.}{2014}]{Bromberg14}
{Bromberg} O.,  {Granot} J.,  {Lyubarsky} Y.,    {Piran} T.,  2014, MNRAS, 443,
  1532, \eprint{1402.4142},
  \adsurl{http://adsabs.harvard.edu/abs/2014MNRAS.443.1532B}

\bibitem[\protect\citeauthoryear{{Bromberg} \& {Levinson}}{{Bromberg} \&
  {Levinson}}{2007}]{BL07}
{Bromberg} O.,  {Levinson} A.,  2007, ApJ, 671, 678, \eprint{0705.2040},
  \adsurl{http://adsabs.harvard.edu/abs/2007ApJ...671..678B}

\bibitem[\protect\citeauthoryear{{Coughlin} \& {Begelman}}{{Coughlin} \&
  {Begelman}}{2014}]{Coughlin14}
{Coughlin} E.~R.,  {Begelman} M.~C.,  2014, ApJ, 781, 82, \eprint{1312.5314},
  \adsurl{http://adsabs.harvard.edu/abs/2014ApJ...781...82C}

\bibitem[\protect\citeauthoryear{{Dixon}}{{Dixon}}{1978}]{Dixon78}
{Dixon} W.~G.,  1978, {Special relativity: the foundation of macroscopic
  physics.}.
Cambridge Univ.~Press,
  \adsurl{http://adsabs.harvard.edu/abs/1978srfm.book.....D}

\bibitem[\protect\citeauthoryear{{Doeleman}, {Fish}, {Schenck}, {Beaudoin},
  {Blundell}, {Bower}, {Broderick}, {Chamberlin}, {Freund}, {Friberg},
  {Gurwell}, {Ho}, {Honma}, {Inoue}, {Krichbaum}, {Lamb}, {Loeb}, {Lonsdale} \&
  {Marrone}}{{Doeleman} et~al.}{2012}]{Doeleman12}
{Doeleman} S.~S.,  {Fish} V.~L.,  {Schenck} D.~E.,  {Beaudoin} C.,  {Blundell}
  R.,  {Bower} G.~C.,  {Broderick} A.~E.,  {Chamberlin} R.,  {Freund} R.,
  {Friberg} P.,  {Gurwell} M.~A.,  {Ho} P.~T.~P.,  {Honma} M.,  {Inoue} M.,
  {Krichbaum} T.~P.,  {Lamb} J.,  {Loeb} A.,  {Lonsdale} C.,    {Marrone}
  D.~P.,  2012, Science, 338, 355, \eprint{1210.6132},
  \adsurl{http://adsabs.harvard.edu/abs/2012Sci...338..355D}

\bibitem[\protect\citeauthoryear{{Eichler}}{{Eichler}}{1982}]{Eichler82}
{Eichler} D.,  1982, ApJ, 263, 571,
  \adsurl{http://adsabs.harvard.edu/abs/1982ApJ...263..571E}

\bibitem[\protect\citeauthoryear{{Eichler}}{{Eichler}}{1993}]{Eichler93}
{Eichler} D.,  1993, ApJ, 419, 111,
  \adsurl{http://adsabs.harvard.edu/abs/1993ApJ...419..111E}

\bibitem[\protect\citeauthoryear{{Georganopoulos} \&
  {Marscher}}{{Georganopoulos} \& {Marscher}}{1998}]{Georganopoulos98}
{Georganopoulos} M.,  {Marscher} A.~P.,  1998, ApJ, 506, 621,
  \eprint{astro-ph/9806170},
  \adsurl{http://adsabs.harvard.edu/abs/1998ApJ...506..621G}

\bibitem[\protect\citeauthoryear{{Goldstein}, {Preece}, {Briggs}, {van der
  Horst}, {McBreen}, {Kouveliotou}, {Connaughton}, {Paciesas}, {Meegan},
  {Bhat}, {Bissaldi}, {Burgess} \& {Chaplin}}{{Goldstein}
  et~al.}{2011}]{Goldstein11}
{Goldstein} A.,  {Preece} R.~D.,  {Briggs} M.~S.,  {van der Horst} A.~J.,
  {McBreen} S.,  {Kouveliotou} C.,  {Connaughton} V.,  {Paciesas} W.~S.,
  {Meegan} C.~A.,  {Bhat} P.~N.,  {Bissaldi} E.,  {Burgess} J.~M.,    {Chaplin}
  V.,  2011, ArXiv e-prints, \eprint{1101.2458},
  \adsurl{http://adsabs.harvard.edu/abs/2011arXiv1101.2458G}

\bibitem[\protect\citeauthoryear{{Jorstad}, {Marscher}, {Lister}, {Stirling},
  {Cawthorne}, {Gear}, {G{\'o}mez}, {Stevens}, {Smith}, {Forster} \&
  {Robson}}{{Jorstad} et~al.}{2005}]{Jorstad05}
{Jorstad} S.~G.,  {Marscher} A.~P.,  {Lister} M.~L.,  {Stirling} A.~M.,
  {Cawthorne} T.~V.,  {Gear} W.~K.,  {G{\'o}mez} J.~L.,  {Stevens} J.~A.,
  {Smith} P.~S.,  {Forster} J.~R.,    {Robson} E.~I.,  2005, AJ, 130, 1418,
  \eprint{arXiv:astro-ph/0502501},
  \adsurl{http://adsabs.harvard.edu/abs/2005AJ....130.1418J}

\bibitem[\protect\citeauthoryear{{Junor}, {Biretta} \& {Livio}}{{Junor}
  et~al.}{1999}]{JunorBiretta99}
{Junor} W.,  {Biretta} J.~A.,    {Livio} M.,  1999, \nat, 401, 891,
  \adsurl{http://adsabs.harvard.edu/abs/1999Natur.401..891J}

\bibitem[\protect\citeauthoryear{{Kohler} \& {Begelman}}{{Kohler} \&
  {Begelman}}{2012}]{Kohler122}
{Kohler} S.,  {Begelman} M.~C.,  2012, MNRAS, 426, 595, \eprint{1208.1261},
  \adsurl{http://adsabs.harvard.edu/abs/2012MNRAS.426..595K}

\bibitem[\protect\citeauthoryear{{Kohler}, {Begelman} \& {Beckwith}}{{Kohler}
  et~al.}{2012}]{Kohler12}
{Kohler} S.,  {Begelman} M.~C.,    {Beckwith} K.,  2012, MNRAS, 422, 2282,
  \eprint{1112.4843},
  \adsurl{http://adsabs.harvard.edu/abs/2012MNRAS.422.2282K}

\bibitem[\protect\citeauthoryear{{Koide}}{{Koide}}{2004}]{Koide04}
{Koide} S.,  2004, ApJ, 606, L45,
  \adsurl{http://adsabs.harvard.edu/abs/2004ApJ...606L..45K}

\bibitem[\protect\citeauthoryear{{Komissarov}}{{Komissarov}}{1994}]{Komissarov94}
{Komissarov} S.~S.,  1994, MNRAS, 269, 394,
  \adsurl{http://adsabs.harvard.edu/abs/1994MNRAS.269..394K}

\bibitem[\protect\citeauthoryear{{Komissarov}}{{Komissarov}}{2011}]{Komissarov11}
{Komissarov} S.~S.,  2011, \memsai, 82, 95, \eprint{1006.2242},
  \adsurl{http://adsabs.harvard.edu/abs/2011MmSAI..82...95K}

\bibitem[\protect\citeauthoryear{{Komissarov}, {Barkov}, {Vlahakis} \&
  {K{\"o}nigl}}{{Komissarov} et~al.}{2007}]{KomBarkVla07}
{Komissarov} S.~S.,  {Barkov} M.~V.,  {Vlahakis} N.,    {K{\"o}nigl} A.,  2007,
  MNRAS, 380, 51, \eprint{arXiv:astro-ph/0703146},
  \adsurl{http://adsabs.harvard.edu/abs/2007MNRAS.380...51K}

\bibitem[\protect\citeauthoryear{{Komissarov} \& {Falle}}{{Komissarov} \&
  {Falle}}{1997}]{Komissarov97}
{Komissarov} S.~S.,  {Falle} S.~A.~E.~G.,  1997, MNRAS, 288, 833,
  \adsurl{http://adsabs.harvard.edu/abs/1997MNRAS.288..833K}

\bibitem[\protect\citeauthoryear{{Komissarov}, {Vlahakis}, {K{\"o}nigl} \&
  {Barkov}}{{Komissarov} et~al.}{2009}]{KomVlaKon09}
{Komissarov} S.~S.,  {Vlahakis} N.,  {K{\"o}nigl} A.,    {Barkov} M.~V.,  2009,
  MNRAS, 394, 1182, \eprint{0811.1467},
  \adsurl{http://adsabs.harvard.edu/abs/2009MNRAS.394.1182K}

\bibitem[\protect\citeauthoryear{{Landau} \& {Lifshitz}}{{Landau} \&
  {Lifshitz}}{1959}]{Landau59}
{Landau} L.~D.,  {Lifshitz} E.~M.,  1959, {Fluid mechanics}.
Butterworth-Heinemann,
  \adsurl{http://adsabs.harvard.edu/abs/1959flme.book.....L}

\bibitem[\protect\citeauthoryear{{Lazzati} \& {Begelman}}{{Lazzati} \&
  {Begelman}}{2005}]{Lazzati05}
{Lazzati} D.,  {Begelman} M.~C.,  2005, ApJ, 629, 903,
  \eprint{astro-ph/0502084},
  \adsurl{http://adsabs.harvard.edu/abs/2005ApJ...629..903L}

\bibitem[\protect\citeauthoryear{{Levinson} \& {Eichler}}{{Levinson} \&
  {Eichler}}{2000}]{Levinson00}
{Levinson} A.,  {Eichler} D.,  2000, Physical Review Letters, 85, 236,
  \eprint{arXiv:astro-ph/0001405},
  \adsurl{http://adsabs.harvard.edu/abs/2000PhRvL..85..236L}

\bibitem[\protect\citeauthoryear{{Lithwick} \& {Sari}}{{Lithwick} \&
  {Sari}}{2001}]{Lithwick01}
{Lithwick} Y.,  {Sari} R.,  2001, ApJ, 555, 540,
  \eprint{arXiv:astro-ph/0011508},
  \adsurl{http://adsabs.harvard.edu/abs/2001ApJ...555..540L}

\bibitem[\protect\citeauthoryear{{Lovelace}}{{Lovelace}}{1976}]{Lovelace76}
{Lovelace} R.~V.~E.,  1976, \nat, 262, 649,
  \adsurl{http://adsabs.harvard.edu/abs/1976Natur.262..649L}

\bibitem[\protect\citeauthoryear{{Lyubarsky}}{{Lyubarsky}}{2009}]{Lyubarsky09}
{Lyubarsky} Y.,  2009, ApJ, 698, 1570, \eprint{0902.3357},
  \adsurl{http://adsabs.harvard.edu/abs/2009ApJ...698.1570L}

\bibitem[\protect\citeauthoryear{{Lyubarsky}}{{Lyubarsky}}{2011}]{Lyubarsky11}
{Lyubarsky} Y.,  2011, \pre, 83, 016302, \eprint{1012.2517},
  \adsurl{http://adsabs.harvard.edu/abs/2011PhRvE..83a6302L}

\bibitem[\protect\citeauthoryear{{McKinney} \& {Blandford}}{{McKinney} \&
  {Blandford}}{2009}]{McKinney09}
{McKinney} J.~C.,  {Blandford} R.~D.,  2009, MNRAS, 394, L126,
  \eprint{0812.1060},
  \adsurl{http://adsabs.harvard.edu/abs/2009MNRAS.394L.126M}

\bibitem[\protect\citeauthoryear{{Meier}}{{Meier}}{2003}]{Meier03}
{Meier} D.,  2003, in {Durouchoux} P.,  {Fuchs} Y.,   {Rodriguez} J.,  eds, New
  Views on Microquasars {The theory of relativistic jet formation in Galactic
  sources: towards a unified model}.
p.~165, \eprint{arXiv:astro-ph/0312047},
  \adsurl{http://adsabs.harvard.edu/abs/2003nvm..conf..165M}

\bibitem[\protect\citeauthoryear{{Nalewajko} \& {Sikora}}{{Nalewajko} \&
  {Sikora}}{2009}]{KN09}
{Nalewajko} K.,  {Sikora} M.,  2009, MNRAS, 392, 1205, \eprint{0810.3912},
  \adsurl{http://adsabs.harvard.edu/abs/2009MNRAS.392.1205N}

\bibitem[\protect\citeauthoryear{{Narayan}, {McKinney} \& {Farmer}}{{Narayan}
  et~al.}{2007}]{Narayan07}
{Narayan} R.,  {McKinney} J.~C.,    {Farmer} A.~J.,  2007, MNRAS, 375, 548,
  \eprint{astro-ph/0610817},
  \adsurl{http://adsabs.harvard.edu/abs/2007MNRAS.375..548N}

\bibitem[\protect\citeauthoryear{{Proga}, {MacFadyen}, {Armitage} \&
  {Begelman}}{{Proga} et~al.}{2003}]{Proga03}
{Proga} D.,  {MacFadyen} A.~I.,  {Armitage} P.~J.,    {Begelman} M.~C.,  2003,
  ApJ, 599, L5, \eprint{astro-ph/0310002},
  \adsurl{http://adsabs.harvard.edu/abs/2003ApJ...599L...5P}

\bibitem[\protect\citeauthoryear{{Sari}, {Piran} \& {Halpern}}{{Sari}
  et~al.}{1999}]{Sari99}
{Sari} R.,  {Piran} T.,    {Halpern} J.~P.,  1999, ApJ, 519, L17,
  \eprint{astro-ph/9903339},
  \adsurl{http://adsabs.harvard.edu/abs/1999ApJ...519L..17S}

\bibitem[\protect\citeauthoryear{{Sikora}, {Begelman} \& {Rees}}{{Sikora}
  et~al.}{1994}]{Sikora94}
{Sikora} M.,  {Begelman} M.~C.,    {Rees} M.~J.,  1994, ApJ, 421, 153,
  \adsurl{http://adsabs.harvard.edu/abs/1994ApJ...421..153S}

\bibitem[\protect\citeauthoryear{{Tchekhovskoy}, {Metzger}, {Giannios} \&
  {Kelley}}{{Tchekhovskoy} et~al.}{2014}]{Tchekhovskoy14}
{Tchekhovskoy} A.,  {Metzger} B.~D.,  {Giannios} D.,    {Kelley} L.~Z.,  2014,
  MNRAS, 437, 2744, \eprint{1301.1982},
  \adsurl{http://adsabs.harvard.edu/abs/2014MNRAS.437.2744T}

\bibitem[\protect\citeauthoryear{{Tchekhovskoy}, {Narayan} \&
  {McKinney}}{{Tchekhovskoy} et~al.}{2010}]{Tchekhovskoy10}
{Tchekhovskoy} A.,  {Narayan} R.,    {McKinney} J.~C.,  2010, \na, 15, 749,
  \eprint{0909.0011},
  \adsurl{http://adsabs.harvard.edu/abs/2010NewA...15..749T}

\bibitem[\protect\citeauthoryear{{Tomimatsu}}{{Tomimatsu}}{1994}]{Tomimatsu94}
{Tomimatsu} A.,  1994, \pasj, 46, 123,
  \adsurl{http://adsabs.harvard.edu/abs/1994PASJ...46..123T}

\bibitem[\protect\citeauthoryear{{Zakamska}, {Begelman} \&
  {Blandford}}{{Zakamska} et~al.}{2008}]{Zakamska08}
{Zakamska} N.~L.,  {Begelman} M.~C.,    {Blandford} R.~D.,  2008, ApJ, 679,
  990, \eprint{0801.1120},
  \adsurl{http://adsabs.harvard.edu/abs/2008ApJ...679..990Z}

\bibitem[\protect\citeauthoryear{{Zauderer}, {Berger}, {Soderberg}, {Loeb},
  {Narayan}, {Frail}, {Petitpas}, {Brunthaler}, {Chornock}, {Carpenter},
  {Pooley} \& {Mooley}}{{Zauderer} et~al.}{2011}]{Zauderer11}
{Zauderer} B.~A.,  {Berger} E.,  {Soderberg} A.~M.,  {Loeb} A.,  {Narayan} R.,
  {Frail} D.~A.,  {Petitpas} G.~R.,  {Brunthaler} A.,  {Chornock} R.,
  {Carpenter} J.~M.,  {Pooley} G.~G.,    {Mooley} K. e.~a.,  2011, \nat, 476,
  425, \eprint{1106.3568},
  \adsurl{http://adsabs.harvard.edu/abs/2011Natur.476..425Z}

\end{thebibliography}

\label{lastpage}

\end{document}